\documentclass[apj]{emulateapj}

\usepackage{apjfonts}

\shorttitle{BH MASS ESTIMATES OF MODERATE-LUMINOSITY AGNS}
\shortauthors{MATSUOKA ET AL.}

\begin{document}

\title{A comparative analysis of virial black-hole mass estimates of moderate-luminosity active galactic nuclei using Subaru/FMOS}

\author{K.~Matsuoka\altaffilmark{1,2}, J.~D.~Silverman\altaffilmark{3}, M.~Schramm\altaffilmark{3}, C.~L.~Steinhardt\altaffilmark{3}, T.~Nagao\altaffilmark{4,5}, J.~Kartaltepe\altaffilmark{6}, D.~B.~Sanders\altaffilmark{7}, E.~Treister\altaffilmark{8}, G.~Hasinger\altaffilmark{7}, M.~Akiyama\altaffilmark{9}, K.~Ohta\altaffilmark{5}, Y.~Ueda\altaffilmark{5}, A.~Bongiorno\altaffilmark{10,11}, W.~N.~Brandt\altaffilmark{12}, M.~Brusa\altaffilmark{13,14,15}, P.~Capak\altaffilmark{16}, F.~Civano\altaffilmark{17}, A.~Comastri\altaffilmark{15}, M.~Elvis\altaffilmark{17}, L.~C.~Ho\altaffilmark{18}, S.~J.~Lilly\altaffilmark{19}, V.~Mainieri\altaffilmark{20}, D.~Masters\altaffilmark{18}, M.~Mignoli\altaffilmark{15}, M.~Salvato\altaffilmark{13}, J.~R.~Trump\altaffilmark{21}, Y.~Taniguchi\altaffilmark{2}, G.~Zamorani\altaffilmark{15}, D.~M.~Alexander\altaffilmark{22}, and Y.~T.~Lin\altaffilmark{23}}

\altaffiltext{1}{Department of Physics and Astronomy, Seoul National University, 599 Gwanak-ro, Gwanak-gu, Seoul 151-742, Republic of Korea}
\altaffiltext{2}{Research Center for Space and Cosmic Evolution, Ehime University, 2-5 Bunkyo-cho, Matsuyama 790-8577, Japan}
\altaffiltext{3}{Kavli Institute for the Physics and Mathematics of the Universe (Kavli IPMU), The University of Tokyo, 5-1-5 Kashiwanoha, Kashiwa 277-8583, Japan}
\altaffiltext{4}{The Hakubi Center for Advanced Research, Kyoto University, Yoshida-Ushinomiya-cho, Sakyo-ku, Kyoto 606-8302, Japan}
\altaffiltext{5}{Department of Astronomy, Kyoto University, Kitashirakawa-Oiwake-cho, Sakyo-ku, Kyoto 606-8502, Japan}
\altaffiltext{6}{National Optical Astronomy Observatory, 950 North Cherry Aveue, Tucson, AZ 85719, USA}
\altaffiltext{7}{Institute for Astronomy, University of Hawaii, 2680 Woodlawn Drive, Honolulu, HI 96822, USA}
\altaffiltext{8}{Departamento de Astronom{\'i}a, Universidad de Concepci{\'o}n, Casilla 160-C, Concepci{\'o}n, Chile}
\altaffiltext{9}{Astronomical Institute, Tohoku University, 6-3 Aramaki, Aoba-ku, Sendai 980-8578, Japan}
\altaffiltext{10}{Max-Planck-Institut f{\"u}r extraterrestrische Physik, Giessenbachstrasse 1, 85748 Garching bei M{\"u}nchen, Germany}
\altaffiltext{11}{INAF -- Osservatorio Astronomico di Roma, Via di Frascati 33, 00040 Monte Porzio Catone, Italy}
\altaffiltext{12}{Department of Astronomy and Astrophysics, The Pennsylvania State University, University Park, PA 16802, USA}
\altaffiltext{13}{Max-Planck-Institut f{\"u}r extraterrestrische Physik, Postfach 1312, 84571 Garching bei M{\"u}nchen, Germany}
\altaffiltext{14}{Dipartimento di Fisica e Astronomia, Universit{\'a} degli Studi di Bologna, viale Berti Pichat 6/2, 40127 Bologna, Italy}
\altaffiltext{15}{INAF -- Osservatorio Astronomico di Bologna, Via Ranzani 1, 40127 Bologna, Italy}
\altaffiltext{16}{NASA/JPL Spitzer Science center, California Institute of Technology, 1200 East California Boulevard, Pasadena, CA 91125, USA}
\altaffiltext{17}{Harvard-Smithsonian Center for Astrophysics, 60 Garden Street, Cambridge, MA 02138, USA}
\altaffiltext{18}{The Observatories of the Carnegie Institution for Science, 813 Santa Barbara Street, Pasadena, CA 91101, USA}
\altaffiltext{19}{Institute for Astronomy, ETH Z{\"u}rich, Wolfgang-Pauli-Strasse 27, 8093 Z{\"u}rich, Switzerland}
\altaffiltext{20}{European Southern Observatory, Karl-Schwarzschild-Strasse 2, 85748 Garching bei M{\"u}nchen, Germany}
\altaffiltext{21}{University of California Observatories/Lick Observatory and Department of Astronomy and Astrophysics, University of California, Santa Cruz, CA 95064, USA}
\altaffiltext{22}{Department of Physics, Durham University, South Road, Durham DH1 3LE, UK}
\altaffiltext{23}{Institute of Astronomy and Astrophysics, Academia Sinica, Taipei 10617, Taiwan}

\email{kenta@astro.snu.ac.kr}

\begin{abstract}
We present an analysis of broad emission lines observed in moderate-luminosity active galactic nuclei (AGNs), typical of those found in X-ray surveys of deep fields, with the aim to test the validity of single-epoch virial black hole mass estimates. We have acquired near-infrared (NIR; $1.0-1.8$ \micron) spectra of AGNs up to $z \sim 1.8$ in the COSMOS and Extended {\it Chandra} Deep Field-South Survey, with the Fiber Multi-Object Spectrograph (FMOS) mounted on the Subaru Telescope. These low-resolution ($R \sim 600$) NIR spectra provide a significant detection of the broad H$\alpha$ emission line that has been shown to be a reliable probe of black hole mass at low redshift. Our sample has existing optical spectroscopy (through programs such as zCOSMOS) which provides a detection of \ion{Mg}{2}, a broad emission line typically used for black hole mass estimation at $z \ga 1$. We carry out a spectral-line fitting procedure using both H$\alpha$ and \ion{Mg}{2} to determine the virial velocity of gas in the broad line region, the monochromatic continuum luminosity at 3000 \AA, and the total H$\alpha$ line luminosity. With a sample of 43 AGNs spanning a range of two decades in luminosity (i.e., $\lambda L_\lambda \sim 10^{44-46}$ ergs s$^{-1}$), we find a tight correlation between the rest-frame ultraviolet and emission-line luminosity with a distribution characterized by $\langle \log(\lambda L_{3000}/L_{\rm H\alpha}) \rangle = 1.52$ and a dispersion $\sigma = 0.16$. There is also a close one-to-one relationship between the FWHM of H$\alpha$ and of \ion{Mg}{2} up to 10000 km s$^{-1}$ with a dispersion of 0.14 in the distribution of the logarithm of their ratios. Both of these then lead to there being very good agreement between H$\alpha$- and \ion{Mg}{2}-based masses over a wide range in black hole mass (i.e., $M_{\rm BH} \sim 10^{7-9} M_\odot$). We do find a small offset in \ion{Mg}{2}-based masses, relative to those based on H$\alpha$, of $+0.17$ dex and a dispersion $\sigma = 0.32$. In general, these results demonstrate that local scaling relations, using \ion{Mg}{2} or H$\alpha$, are applicable for AGN at moderate luminosities and up to $z \sim 2$.
\end{abstract}

\keywords{black hole physics --- galaxies: active --- quasars: emission lines}

\section{Introduction}

While the study of galaxy evolution has made important strides in recent years by being able to weigh individual galaxies (i.e., determine a mass), the field of quasar research is grappling with the issue of how to measure accurately the masses of a supermassive black holes (SMBHs) for the distant quasar population. Here the challenge is greater due to the fact that the sphere of influence of a SMBH can only be resolved for a limited sample of nearby galaxies whereas the dynamical mass of a galaxy can easily be measured due to its large spatial extent. A significant leap forward in our ability to both accurately and efficiently measure the masses of SMBHs, $M_{\rm BH}$, at all redshifts will likely lead to new insights on questions such as how are black holes fueled, what it is the connection with its host galaxy, and how do SMBHs evolve within a cosmological framework.

Spectroscopy enables us to probe the kinematics of ionized gas within the vicinity of a SMBH in distant active galactic nuclei (AGNs) and luminous quasars to infer their black hole masses. Traditionally, emission lines (e.g., \ion{C}{4}, \ion{Mg}{2}, H$\beta$, and H$\alpha$) detected in the optical and velocity-broadened between $2000-20000$ km s$^{-1}$ are used to probe the gravitational potential well of a SMBH. This lower limit on the velocity width has been set somewhat arbitrarily since there exists a well-known population of both type 1 AGNs having narrower line widths \citep[i.e., NLS1;][]{1987ApJ...323..108O} and those with intermediate-mass black holes \citep{2004ApJ...610..722G,2007ApJ...670...92G}.

There are methods to determine the luminosity-weighted radial distance between the broad-line region (BLR) and central source, $R_{\rm BLR}$, for AGN ($z \la 0.4$) through reverberation-mapping campaigns \citep[e.g.,][]{1972ApJ...171..467B,1993PASP..105..247P,2009ApJ...705..199B} based on Balmer lines. Even with the complex nature of the BLR, this characteristic radius is tightly correlated with its luminosity \citep{2000ApJ...533..631K,2005ApJ...629...61K}, thus providing a means to infer such a distance to the BLR in large quasar samples based solely on luminosity. Then coupled with velocity information provides a viral mass estimate based on a single-epoch spectrum. Such techniques have been applied to large quasar samples most notably the Sloan Digital Sky Survey (SDSS). A number of studies \citep[e.g.,][]{2009ApJ...699..800V,2010MNRAS.402.2637S,2011ApJS..194...45S} clearly demonstrate that such samples effectively probe SMBHs above $M_{\rm BH} \ga 10^9 M_\odot$ at $z \sim 2$ (an epoch of maximal black hole activity) due to the wide area coverage and shallow depth. It is important to keep in mind that these black hole masses are based on calibrations using lower luminosity AGNs at low redshift; their application to luminous quasars at high redshift is not well solidified with reverberation mapping \citep{2007ApJ...659..997K}.

Deep surveys, such as COSMOS, GOODS and AEGIS, are effective probes of black hole accretion at lower masses. Given that the black hole mass function is steeply declining at $\log M_{\rm BH} \ga 9$ \citep{2012ApJ...746..169S}, studies of the global population with SDSS are susceptible to large uncertainties when extrapolated to lower masses \citep{2012arXiv1209.0477K}. While noble attempts have been made to characterize the low-mass end \citep[$\log M_{\rm BH} \la 9$;][]{2006MNRAS.365..134T,2007ApJ...654..731H,2008MNRAS.388.1011M}, these studies have been based on luminosity and do not consider virial velocities. These deep survey fields have considerable X-ray coverage that can be utilized to select AGNs that mitigate biases incurred by host galaxy dilution and obscuration. Such selection then has the potential to effectively probe the lower luminosity AGN population that may be powered by lower mass black holes or those accreting at sub-Eddington rates. Followup optical spectroscopic observations are enabling single-epoch virial black hole mass estimates (down to $10^7 M_\odot$) based on the properties of their broad emission lines and continuum luminosity \citep{2009ApJ...700...49T,2010ApJ...708..137M}.

Mass estimates for these higher redshift AGNs, which actually constitute the majority of the population in deep surveys such as COSMOS, rely on \ion{Mg}{2} or \ion{C}{4} since the H$\beta$ line (used to calibrate recipes based on local samples) is no longer available in the optical window at $z \ga 0.8$. While the assumption that the \ion{Mg}{2} line is produced from the same physical region as H$\beta$ may hold \citep{2002MNRAS.337..109M,2008ApJ...680..169S}, there are studies that indicate that the physics of the broad-line region is not so simple \citep{2009ApJ...707.1334W,2012MNRAS.427.3081T}, especially for the most luminous quasars \citep{2009A&A...495...83M,2011arXiv1109.1554S,2013arXiv1301.0520M} that can have significant outflows possibly in response to a more intense radiation field. Furthermore, it is non-trivial to disentangle the broad \ion{Mg}{2} line from Fe emission that sits at its base, especially for lower mass black holes and maybe even those at the high mass end.

We present first results of a near-infrared spectroscopic survey of broad-line AGNs (BLAGNs) primarily in the COSMOS field using the Fiber Multi-Object Spectrograph (FMOS) mounted on the Subaru Telescope. With FMOS, we now have the capability to simultaneously acquire near-infrared spectra of $\sim 200$ targets over a field of view of 0.19 square degrees. In this study, we report on the comparison between the H$\alpha$ emission line profile, detected in the near-infrared, with that of \ion{Mg}{2} present in previously available optical spectra. We aim to establish how effective recipes (established locally) are to measure single-epoch black hole masses out to $z \sim 1.8$ and for BLAGNs of lower luminosity (i.e., lower black hole mass), as compared to those found in the SDSS. Our sample is supplemented with AGNs from the Extended {\it Chandra} Deep Field-South (ECDF-S) Survey that reach fainter depths, in X-rays, than the {\it Chandra}/COSMOS survey and improve our statistics at lower black hole masses. In Section~\ref{xxfmos}, we fully describe the FMOS observations including target selection, data reduction, and success rate with respect to detecting the H$\alpha$ emission line. We describe our method for fitting broad emission lines in Section~\ref{xxxfit}. Our results are described in Section~\ref{result}. Throughout this work, we assume $H_0 = 70 $ km s$^{-1}$ Mpc$^{-1}$, $\Omega_\Lambda = 0.7$, $\Omega_{\rm M} = 0.3$, and AB magnitudes.

\section{Near-infrared spectroscopy with Subaru/FMOS}\label{xxfmos}

The capability of FMOS \citep{2010PASJ...62.1135K} to simultaneously acquire near-infrared spectra for a large number of objects over a wide field offers great potential for studies of galaxies \citep{2012PASJ...64...60Y,2012MNRAS.426.1782R} and AGNs \citep{2012ApJ...761..143N} at high redshift. Over a circular region of 30$\arcmin$ in diameter, it is possible to place up to 400 fibers, each with a 1\farcs2 aperture, across the field. To detect emission lines in AGNs over a wide range in redshift, we elect to use the low-resolution mode that effectively covers two wavelength intervals of $1.05-1.34$ \micron\ (J band) and $1.43-1.77$ \micron\ (H band) simultaneously. The spectral resolution is $\lambda / \Delta \lambda \approx 600$, thus a velocity resolution FWHM $\sim 500$ km s$^{-1}$ at $\lambda = 1.5$ \micron, suitable for the study of broad emission lines of AGN. Unfortunately, this mode requires an additional optical element (i.e., VPH grating) that reduces the total throughput to $\sim 4$\% at 1.3 \micron\ and impacts the limiting depth reachable in a few hours of integration time.

Accurate removal of the bright sky background when observing from the ground with near-infrared detectors remains the primary challenge. With FMOS, an OH-airglow suppression filter \citep{1993PASP..105..940M,2001PASJ...53..355I} is built into the system that significantly reduces the intensity of strong atmospheric emission lines that usually plague the J band and H band. Equally important, there is the capability (cross-beam-switching; CBS hereafter) to dither targets between fiber pairs effectively measuring the sky background spatially close to individual objects and through the same fibers as the science targets. In this mode, two sequential observations are taken by offsetting the telescope by 60$\arcsec$ while keeping the target within one of the two fibers. The trade off is that only 200 fibers are available in this CBS mode. We refer the reader to \citet{2010PASJ...62.1135K} for full details of the instrument and its performance.

Below, we briefly describe our observing program using FMOS including the selection of type 1 AGNs, data reduction, and success rate with respect to the detection of broad emission lines. Complete details of our program will be presented in Silverman et al. (in prep.) along with the full catalog of emission-line properties of broad-line AGNs in the COSMOS and ECDF-S.

\subsection{Target selection}

Our primary selection of AGNs is based on their having X-ray emission detected by {\it Chandra} \citep{2009ApJS..184..158E,2012ApJS..201...30C} within the central square degree of COSMOS (hereafter C-COSMOS). The high surface density of AGNs ($\sim 2000$ per deg$^{2}$) at the limiting depths ($f_{0.5-2.0 \ {\rm keV}} > 2 \times 10^{-16}$ ergs cm$^{-2}$ s$^{-1}$) of C-COSMOS field ensures that we make adequate use of the multiplex capabilities of FMOS. In addition, two FMOS pointings were observed further out from the center of the COSMOS field thus we relied upon the catalog of optical and near-infrared counterparts to {\it XMM-Newton} sources \citep{2010ApJ...716..348B} for AGN selection.

We further require that optical spectroscopy \citep{2007ApJS..172..383T,2009ApJS..184..218L} is available for each source that yields a reliable redshift and detection of at least a single broad (FWHM $> 2000$ km s$^{-1}$) emission line, namely \ion{Mg}{2} in many cases. We then specifically targeted those with spectroscopic redshifts that allows us to detect either H$\beta$ ($1.2 < z < 1.7$ and $1.9 < z < 2.6$), H$\alpha$ ($0.6 < z < 1.0$ and $1.2 < z < 1.7$), or \ion{Mg}{2} ($2.8 < z < 3.8$ and $4.1 < z < 5.3$) in the observed FMOS spectral windows in low-resolution mode, i.e., $1.05-1.34$ \micron\ and $1.43-1.77$ \micron. Fibers are assigned to BLAGNs (for which we can detect emission lines of interest) with a limiting magnitude of $J_{\rm AB} = 23$. Those at $J_{\rm AB} < 21.5$ are given higher priority to ensure that this sample (of lower density on the sky) is well represented in the final catalog; this also effectively improves upon our success rate of detection both continuum and line emission. Due to the sensitivity of FMOS and the low number density of AGNs at $z > 3$, our sample has very few detections of \ion{Mg}{2} in the near infrared.

In addition, we have acquired FMOS observations of X-ray selected AGNs in the ECDF-S \citep{2005ApJS..161...21L} survey with the inclusion of those that are only detected in the deeper 2$-$4 Ms data \citep{2010ApJS..187..560L,2011ApJS..195...10X} in the central region that covers GOODS. This deeper X-ray field offers the potential to extend the dynamic range of our study in terms of black hole mass and Eddington ratio. We specifically select AGNs, as mentioned above, based on their optical properties determined through deep spectroscopic campaigns \citep{2004ApJS..155..271S,2010ApJS..191..124S}. As with the COSMOS sample, we place higher priority on the brighter AGNs ($R_{\rm AB} < 22$) while targeting the fainter cases with lower priority.

\subsection{Observations}

We have acquired near-infrared spectra with Subaru/FMOS of broad-line AGNs from the COSMOS and ECDF-S surveys. The majority of the data was obtained during open use time through NAOJ over three nights in December 2010 (ID; S10B-108) and two nights in December 2011 (ID; S11B-098). Additional targets were observed during other programs being carried out in the COSMOS field through the University of Hawaii in S10B$-$S11A. Weather conditions were acceptable although clouds, mainly cirrus, reduced our observing efficiency. The typical seeing was $\sim 1\arcsec$ with considerable variation across the nights.

We elected to use the CBS mode while taking two sequential exposures each of 15 minutes for each position (namely A and B positions hereafter). These pairs of exposure were repeated multiple times in order to reach an effective total integration time of $2.0-3.5$ hours on-source. Some time is lost to refocusing and repositioning fibers at regular intervals during the full observation. In the early data, only one spectrograph (IRS1) was available thus $\sim 100$ fibers were available for science targets.

\begin{figure}
\epsscale{1.1}
\plotone{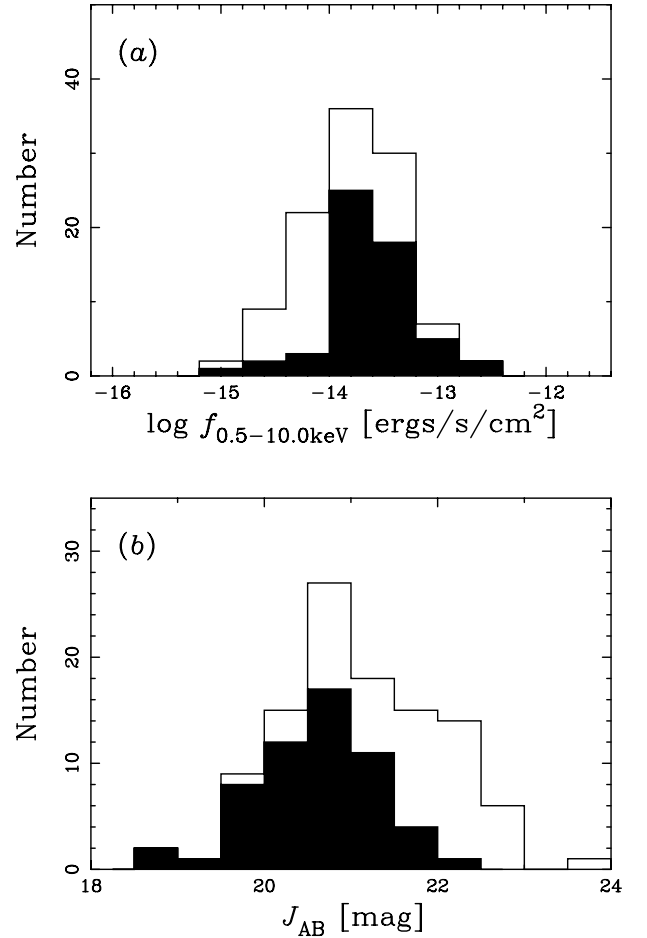}
\caption{Selection and completeness of the COSMOS AGNs with FMOS spectra. Each panel shows the distribution of the observed sample (open histogram) and those that yield black hole mass estimates based on the H$\alpha$ emission line (filled black histogram) for ($a$) X-ray $0.5-10.0$ keV flux and ($b$) J-band magnitude.}
\label{sample}
\end{figure}

\subsection{Data Reduction}

We use the publicly available software FIBRE-pac \citep[FMOS Image-Based REduction package;][]{2012PASJ...64...59I}. The reduction routines are based on IRAF tasks although several steps are processed by additional tools written by the FMOS instrument team. Since our observation are carried out using an ABAB nodding pattern in CBS mode of the telescope, an effective sky subtraction (A$-$B) can be performed using the two different sky images: A$_n -$ B$_{n-1}$ and A$_n -$ B$_{n+1}$ taken before and after the $n$-th exposure. After the initial background subtraction, a cross talk signal is removed by subtracting 0.15\% for IRS1 and 1\% for IRS2 from each quadrant. The difference in the bias between the quadrants is corrected to make the average over each quadrant equal. We further apply a flat field correction using a dome lamp exposure. Bad pixels are masked throughout the reduction procedure. Additional steps include the distortion correction and the removal of residual airglow lines. This procedure is carried out for both positions A and B. Individual frames are combined into an averaged image and an associated noise image. Finally, the wavelength calibration is carried out based on a reference image of a Th-Ar emission spectrum. Individual one-dimensional science and error spectra are extracted both to be used for the fitting of emission line profiles.

We perform a first attempt at flux calibration by using spectra of bright stars, usually $1-2$ per spectrograph. A single stellar spectrum for each spectrograph is chosen to apply a correction based on the spectroscopic magnitude and photometry from 2MASS. An improvement of the absolute flux level is required to account for aperture effects. Therefore, we scale the flux of each FMOS spectra of our AGNs to match the deep infrared photometry using total magnitudes available from the UltraVISTA survey \citep{2012A&A...544A.156M} available over the COSMOS field. While scale factors can reach as high as $\sim 4$, the median value is 1.64.

\subsection{Sample characteristics and completeness}

We have observed over 100 type 1 AGNs in the combined COSMOS and ECDF-S fields to date. In Figure~\ref{sample}, we show the X-ray flux and NIR magnitude distribution of the 108 AGNs (originally identified as {\it Chandra} X-ray sources) in the COSMOS field that have $0.6 < z < 1.8$, a redshift interval where we are capable of detecting H$\alpha$ in the FMOS spectroscopic window. The distributions are shown for both the observed objects and the 56 having a significant detection of a broad H$\alpha$ emission line (at the expected wavelength) that subsequently yielded a black hole mass estimate.

\begin{figure}
\epsscale{1.1}
\plotone{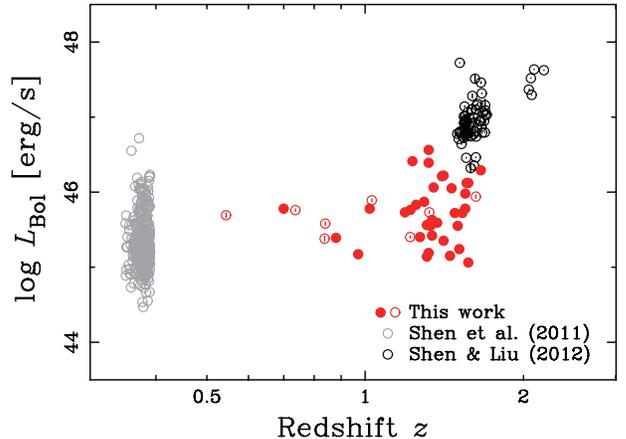}
\caption{Bolometric luminosity as a function of redshift for AGNs in COSMOS (red filled circles), ECDF-S (red open circles), and SDSS (grey and black symbols).}
\label{lbol_z}
\end{figure}

Prior to our observations, we had no empirical assessment of the performance of FMOS thus AGN were targeted to faint infrared magnitudes, now understood not to be feasible using the low-resolution mode for the faintest objects ($J_{\rm AB} \sim 23$). As shown in Figure~\ref{sample}($b$), we have a reasonable level of success (71\%) with the detection of broad emission lines at brighter magnitudes ($J_{\rm AB} < 21.5$). Unfortunately, the success rate is significantly lower at fainter magnitudes thus dropping to 50\% for the entire sample with $J_{\rm AB} < 23$. It is worth highlighting that a survey depth of $J_{\rm AB} < 21.5$ is about two magnitudes fainter than current near-infrared spectroscopic observations of SDSS quasars \citep{2012ApJ...753..125S} at similar redshifts. In Figure~\ref{lbol_z}, we demonstrate this by plotting the bolometric luminosity \citep[based on $L_{3000}$ and a bolometric correction of 5.15;][]{2011ApJS..194...45S} of our AGN compared to those from SDSS surveys.

\citet{2005ApJ...630..122G} describe in detail the benefits of using H$\alpha$ for black hole mass measurements. In particular, the H$\alpha$ emission line is stronger ($\sim 3 \times$) thus more easily detected as compared to H$\beta$. This is clearly evident from our observations. We successfully detect a broad H$\alpha$ line in $\sim 50$\% of the cases (as mentioned above) while the detection of H$\beta$ is very low (17\%). Even so, we do detect H$\beta$ in a fair number of cases up to $z \sim 2.6$ that will be presented in the full emission-line catalog (Silverman et al. in prep.). Any future FMOS campaign designed to detect the H$\beta$ emission line (both broad and narrow) should increase the exposure time significantly and/or use the high-resolution mode that has roughly three times higher throughput in the H band as compared to low-resolution mode \citep[see Figure~19 of][]{2010PASJ...62.1135K}.

\begin{figure}
\epsscale{1.1}
\plotone{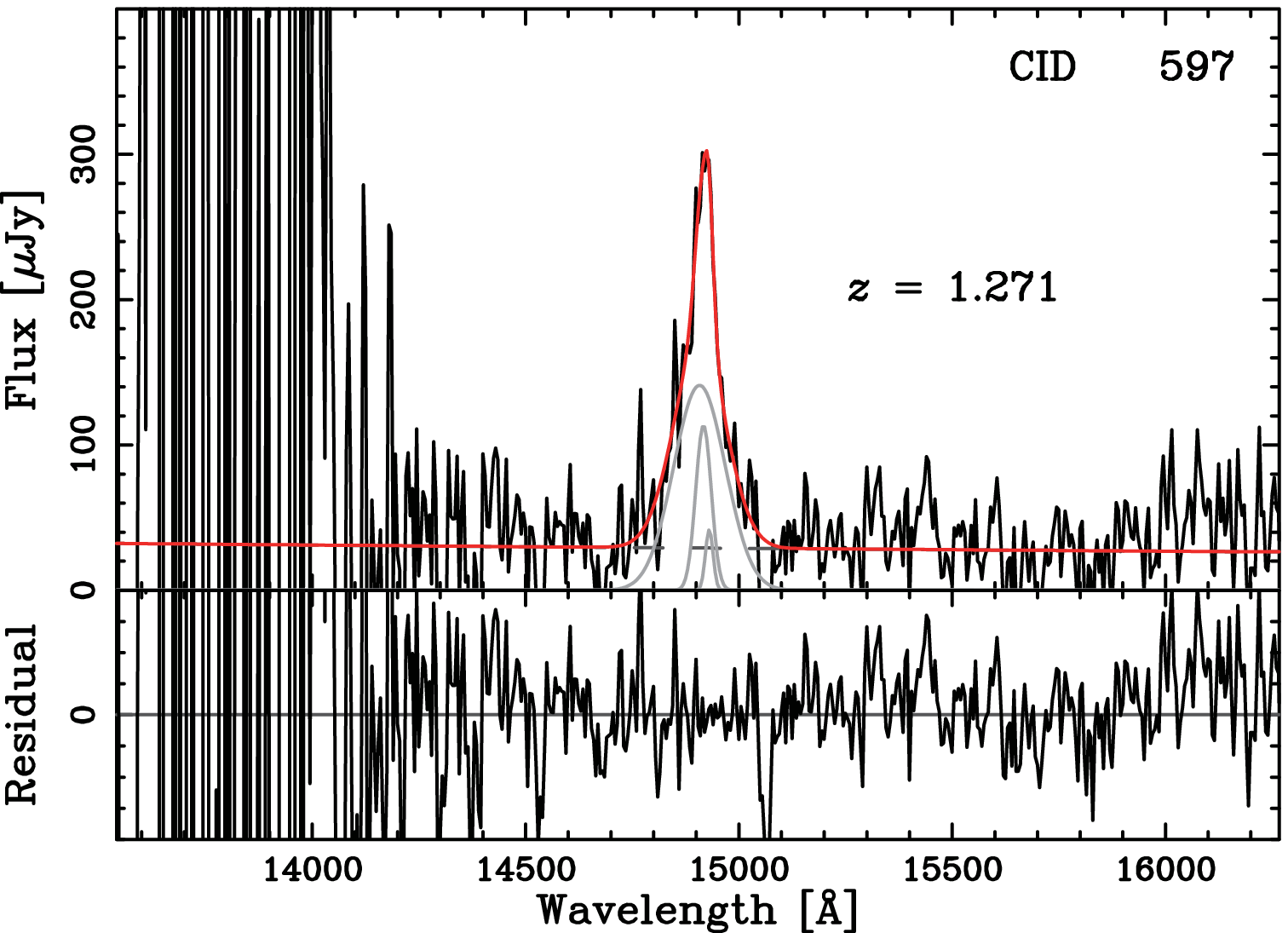}
\vspace{3.0mm}
\plotone{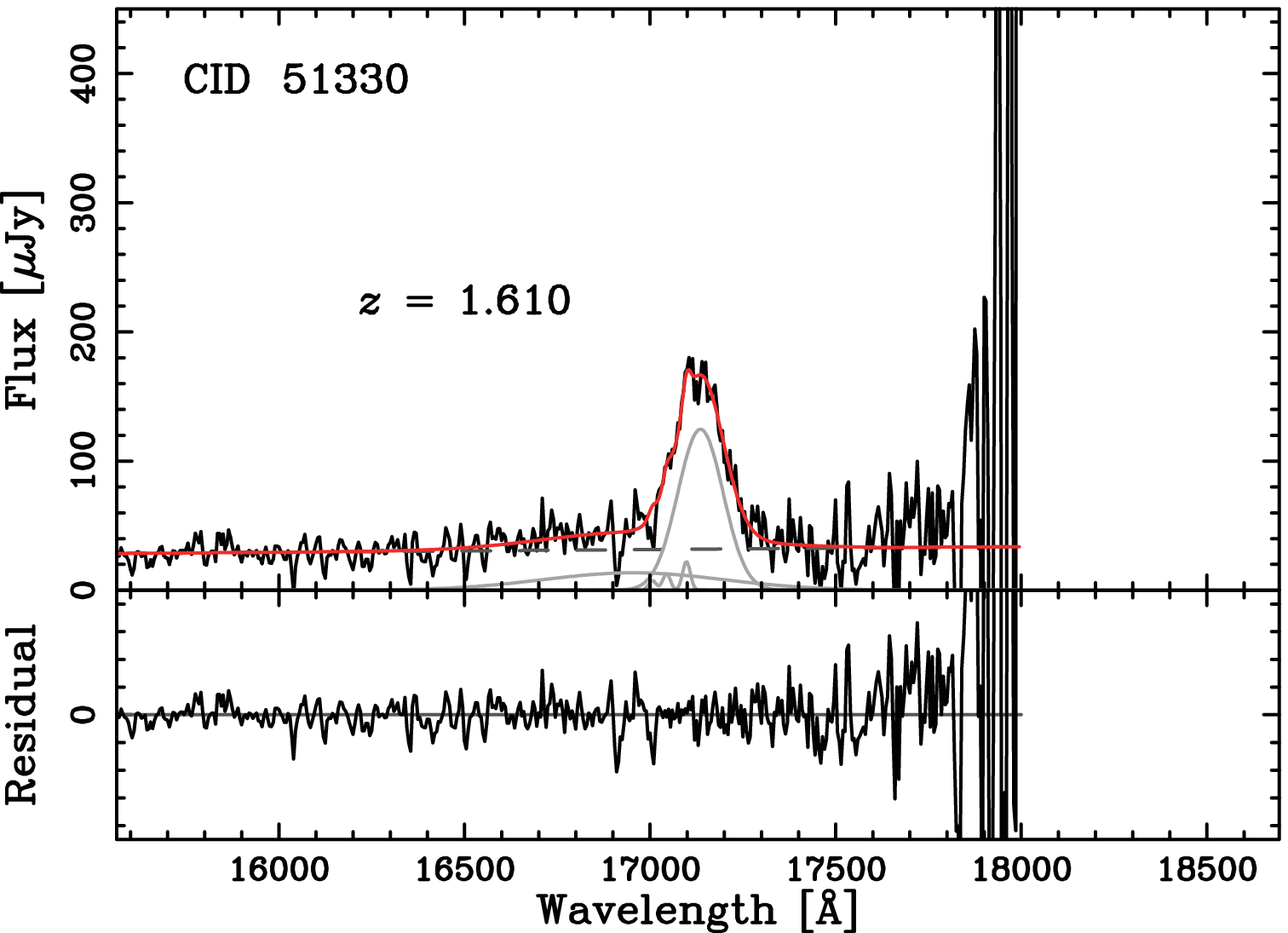}
\vspace{3.0mm}
\plotone{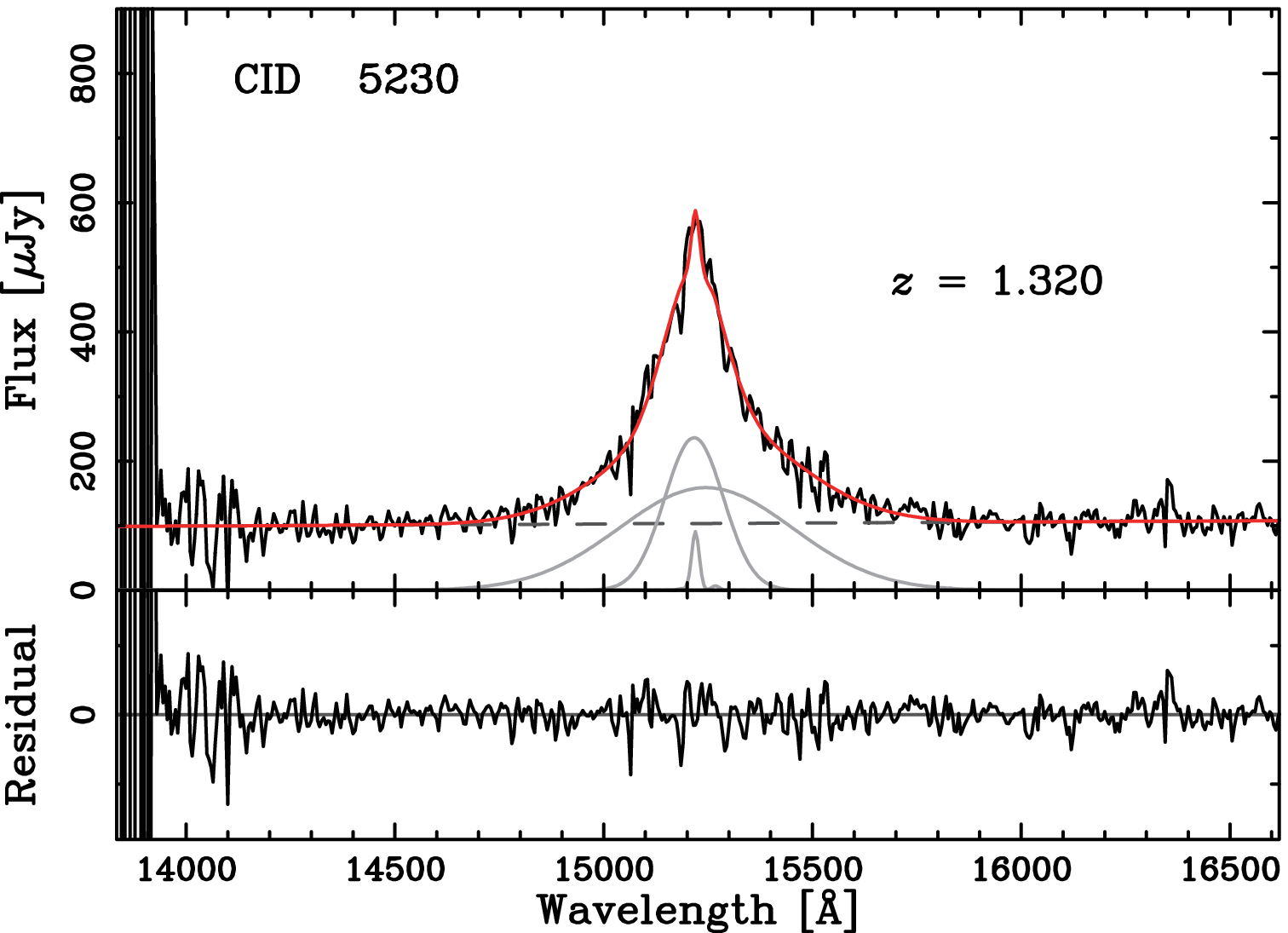}
\caption{Examples of our fitting routine for the H$\alpha$ line detected in COSMOS AGN. In the upper portion of each panel, the observed spectrum is shown in black with the best fit in red. The continuum (gray dashed line) and individual emission-line components (grey gaussian curves) are also indicated respectively. In lower panels, the residuals are shown with same scale as that of upper panel.}
\label{haexam}
\end{figure}

\section{Spectral line fitting}\label{xxxfit}

We measure the properties of broad emission lines present in optical and near-infrared spectra to derive single-epoch black hole mass estimates. For the emission lines of interest here (i.e., \ion{Mg}{2} and H$\alpha$), we specifically measure the full width at half maximum (FHWM), total luminosity of the emission line in the case of H$\alpha$, and continuum luminosity at 3000 \AA. Due to the moderate luminosities of the AGN sample, there can be a non-negligible host galaxy contribution that impacts the estimate of the AGN continuum at redder wavelengths; therefore, we chose to use the H$\alpha$ line flux rather than the continuum luminosity. Fortunately, the multi-wavelength photometry of the COSMOS, including the HST imaging, enables us to determine the level of such contamination that will be fully assessed in a future study. Emission lines are fit using a procedure as outlined below that enables us to characterize the line shape for even those that have a considerable level of noise.

Our fitting procedure of the continuum and line emission utilizes {\tt MPFITFUN}, a Levenberg-Marquardt least squares minimization algorithm as available within the IDL environment. Even though this routine has well-known computational issues, this algorithm is widely used due to its ease of use and fast execution time. The routine returns best-fit parameters and their errors as well as measure of the goodness of the overall fit. We further describe the individual components required to successfully extract a parameterization of the broad component used in determining virial masses. It is worth recognizing that each line has its own advantages and disadvantages that need to be considered carefully especially when fitting data of moderate signal-to-noise ratio (S/N). A final inspection of each fit by eye is performed to remove obvious cases where a broad component is not adequately determined almost exclusively due to spectra having low S/N.

\begin{figure}
\epsscale{1.1}
\plotone{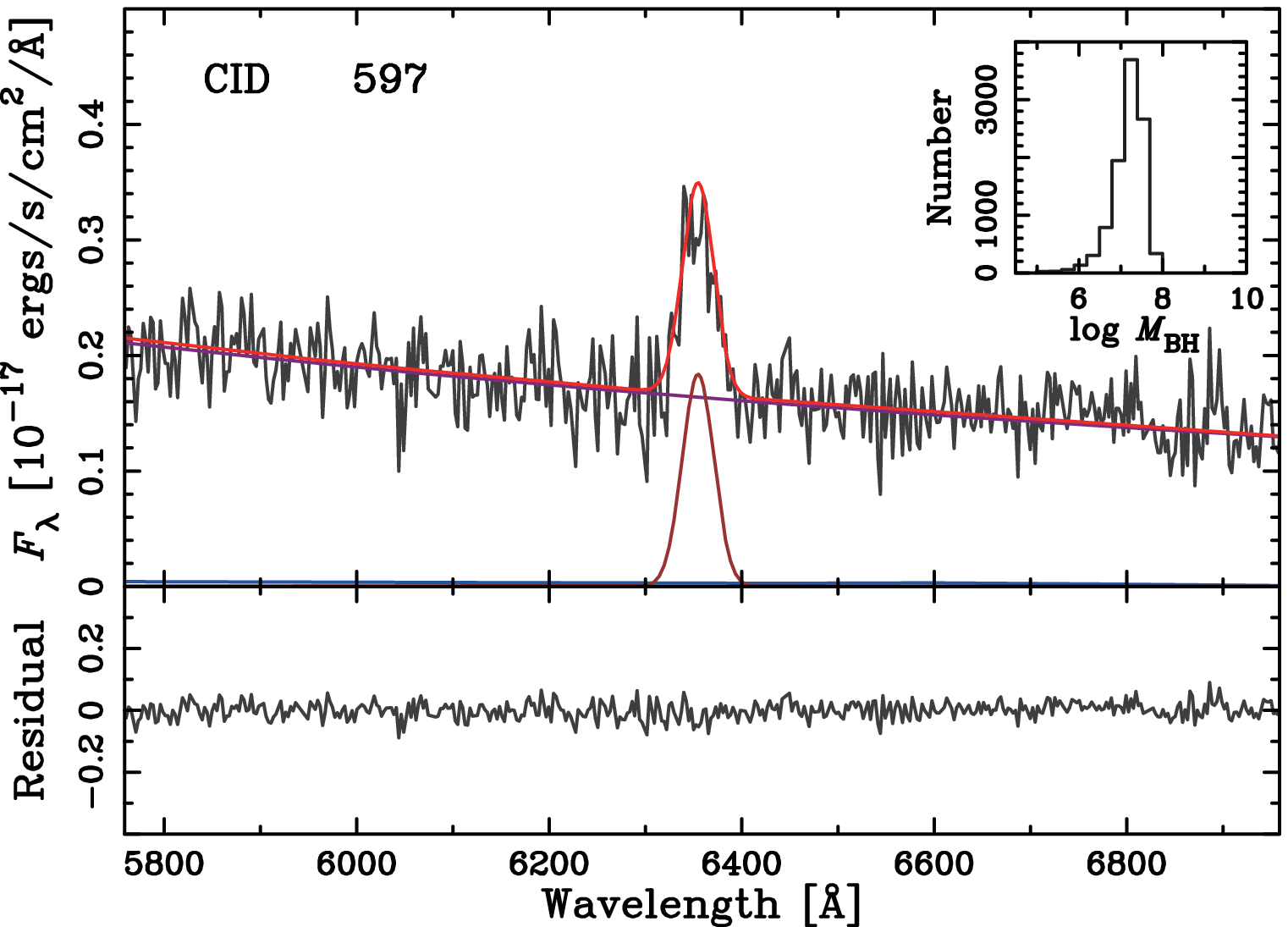}
\vspace{3.0mm}
\plotone{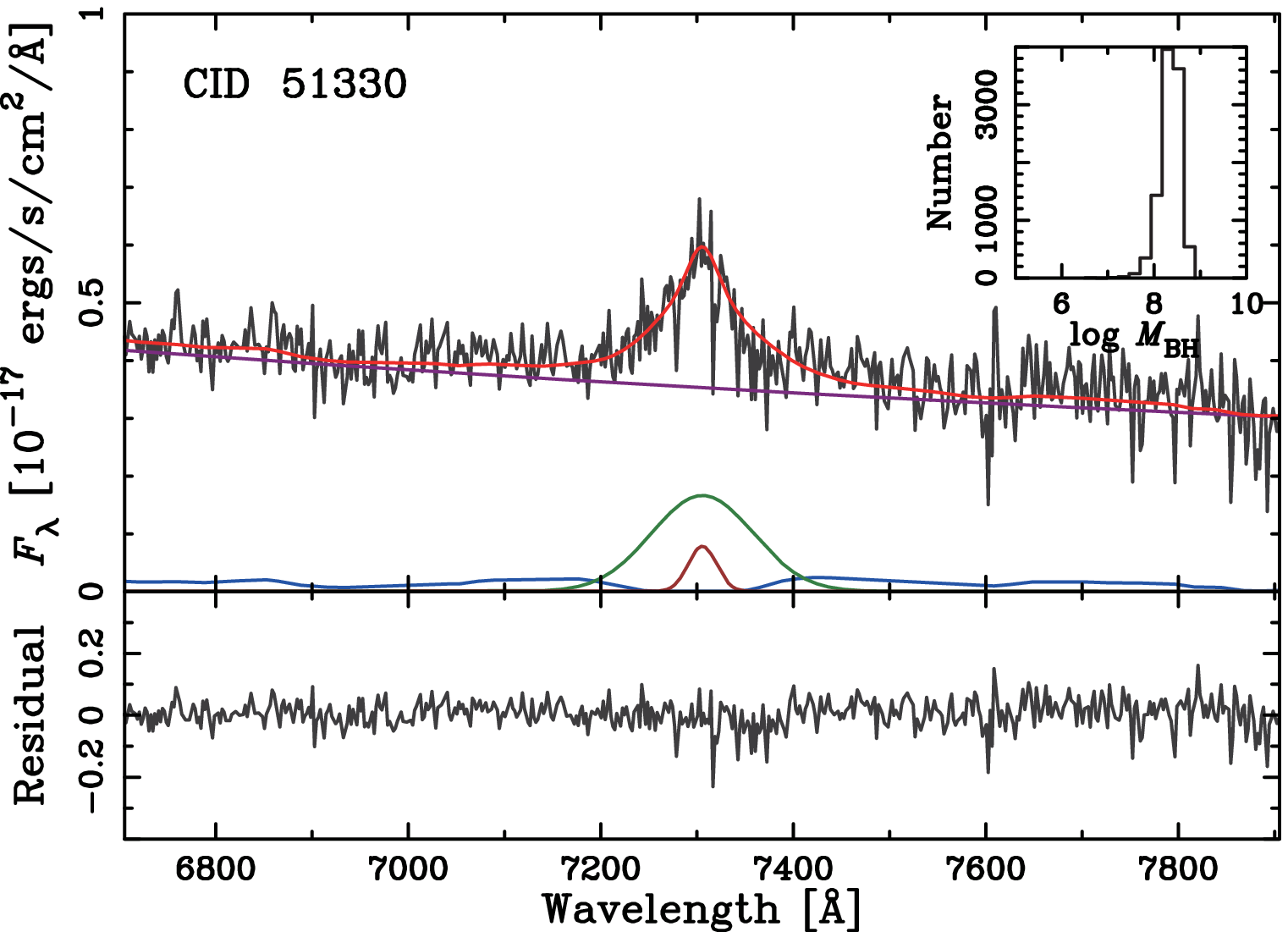}
\vspace{3.0mm}
\plotone{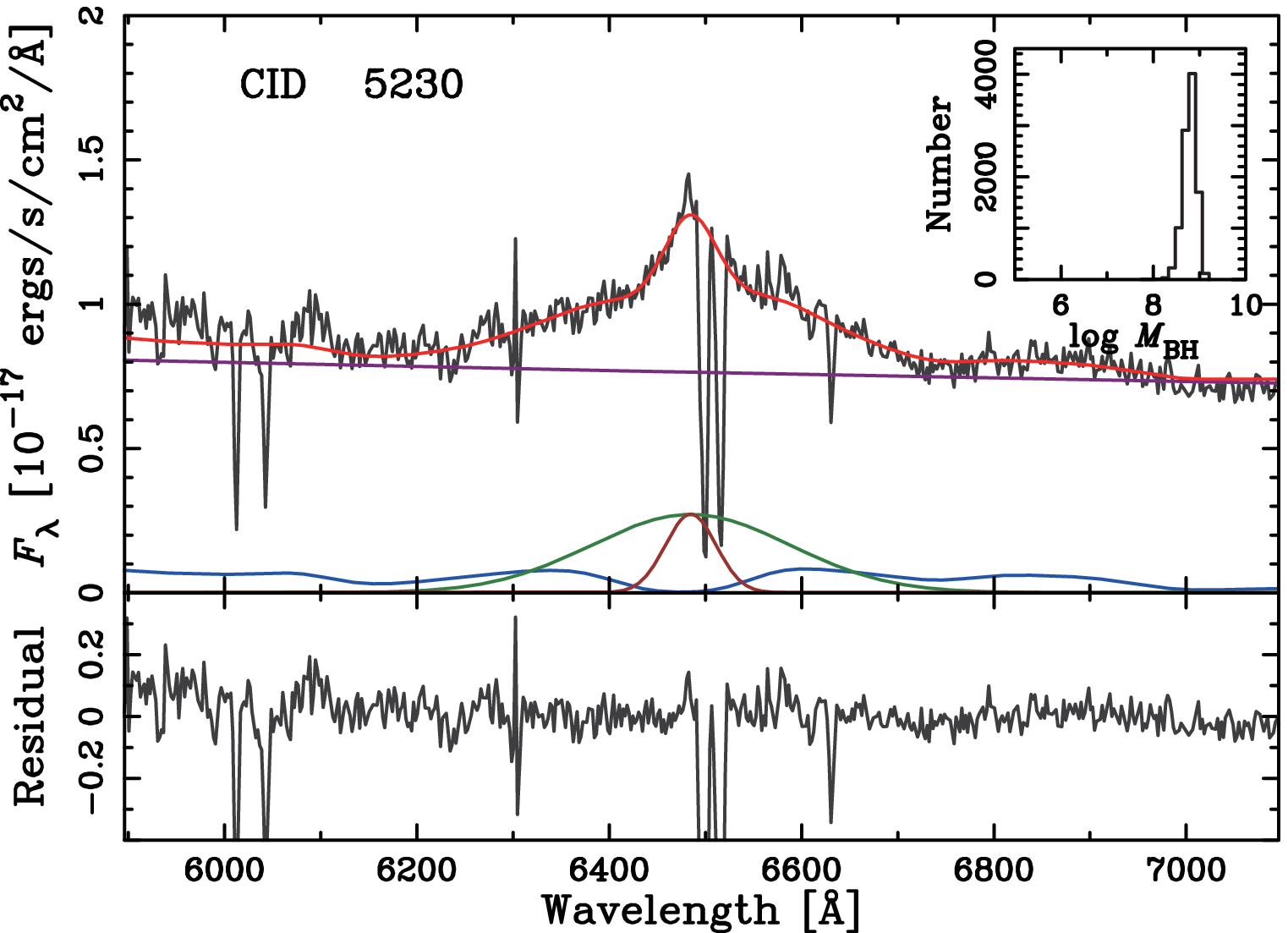}
\caption{Examples of our fits for the \ion{Mg}{2} line for the same AGN as shown in Figure~\ref{haexam}. The top panel shows the spectral range around \ion{Mg}{2}. The best fit model is indicated as a red solid line. The different components are as indicated: Fe-emission (blue), pseudo continuum (purple), and Gaussian components (green and brown). The residual is shown in the lower panel. In the upper right corner for each panel, we show the black hole mass distribution computed from our Monte Carlo tests based on the uncertainties of the line width and continuum luminosity measurements.}
\label{mgexam}
\end{figure}

\subsection{H$\alpha$}

We perform a fit to the H$\alpha$ emission line (if detected within the FMOS spectral window) in order to measure line width and integrated emission-line luminosity. Based on the spectroscopic redshift as determined from optical spectroscopy, we select the spectrum at rest wavelengths centered on the emission line and spanning a range that enables an accurate determination of the continuum characterized by a power law, $f_\lambda \propto \lambda^{-\alpha}$.

We employ multiple Gaussian components to describe the line profile. It is common practice to make such an assumption on the intrinsic shape of individual components, even though it has been demonstrated that broad-emission lines in AGN are not necessarily of such a shape \citep[e.g.,][]{2006A&A...456...75C}. The H$\alpha$ line is fit with two or three Gaussians (including a narrow component) and the [\ion{N}{2}]$\lambda$6548,6684 lines with a pair of Gaussians. The ratio of the [\ion{N}{2}] lines is fixed at the laboratory value of 2.96. The narrow width of the [\ion{N}{2}] lines is fixed to match the narrow component of H$\alpha$. The width of the narrow components is limited to $420-800$ km s$^{-1}$ (a range not corrected for intrinsic dispersion). The velocity profile of the broad components is characterized by the FWHM, measured using either one gaussian or the sum of two gaussians. We then correct the velocity width for the effect of instrumental dispersion to achieve an intrinsic profile width. The H$\alpha$ luminosity discussed throughout this work is the sum of the broad components.

There are cases for which the fitting routine returns a solution with the width of the narrow component pegged at the upper bound of 800 km s$^{-1}$. It is worth highlighting that this minimization routine stops the fitting procedure when a parameters hit a limit thus the returned values are not the true best-fit values. For these, we inspect all fits by eye and decide whether such an additional broad component is real. For many cases, we can use the [\ion{O}{3}]$\lambda$5007 line profile, within the FMOS spectral window, to determine whether such a fit to the narrow line complex is accurate. In addition, we can use the available optical spectra for such comparisons. When the level of significance of the narrow line is negligible, we rerun the fitting routine and fix the narrow line width to the spectral resolution of FMOS, $\sim 420$ km s$^{-1}$. In Figure~\ref{haexam}, we show three examples of our fits to the H$\alpha$ emission line that span a range of line properties.

\subsection{\ion{Mg}{2}}

We fit the \ion{Mg}{2} emission line, as done in \citet{2012arXiv1212.2999S}, observed in optical spectra primarily from zCOSMOS \citep{2009ApJS..184..218L}, Magellan/IMACS \citep{2007ApJS..172..383T}, and SDSS \citep{2000AJ....120.1579Y}. The emission line is modeled by a combination of one or two broad Gaussian functions to best characterize the line shape. We first remove the continuum (before attempting to deal with the emission lines) by fitting the emission in a window surrounding the line. As with H$\alpha$, a power-law function is chosen to best characterize the featureless, non-stellar light attributed to an accretion disk. We further include a broad Fe emission component based on an empirical template \citep{2001ApJS..134....1V} that is convolved by a Gaussian of variable width and straddles the base of the \ion{Mg}{2} emission line. A least square minimization is implemented to determine the best-fit parameters. When possible, we optimize residuals of the fits on a case-by-case basis by trying to minimize the number of components.  Absorption features are either masked out or interpolated across. The fit returns two parameters required for black hole mass estimates: FWHM and monochromatic luminosity at 3000 \AA. Examples of our fits to the \ion{Mg}{2} line are presented in Figure~\ref{mgexam}.

\section{Results}\label{result}

We can determine how closely the parameters (i.e., luminosity and FWHM) required to estimate single-epoch black hole masses agree between the H$\alpha$ and \ion{Mg}{2} emission lines. Any systematic offset or inherent scatter may only add additional uncertainty to the derived masses. We essentially want to establish whether or not the kinematics of the BLR is consistent with photoionized gas in virial motion around the SMBH. We provide all measurements and derived masses in Table~\ref{catalg}.

\subsection{Luminosities}

\begin{figure}
\epsscale{1.1}
\plotone{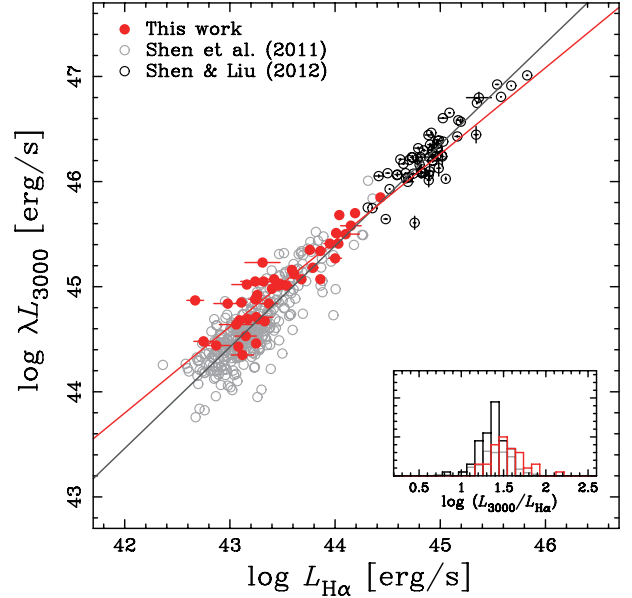}
\caption{Comparison between the monochromatic luminosity at 3000 \AA\ and the H$\alpha$ emission-line luminosity. AGNs in our sample are shown as red circles. The open circles mark published observations of SDSS quasars (grey circles: $0.35 < z < 0.40$; black circles: $1.5 < z < 2.2$). A linear fit to both our data only (red line) and the joint sample (black line) is given. The distribution of $\log (\lambda L_{3000} / L_{\rm H\alpha})$ as a histogram is shown in the sub panel. The colors of their histograms are same as the symbols. Errors represent a 1$\sigma$ level of uncertainty.}
\label{lumlum}
\end{figure}

Our first concern is to determine whether the H$\alpha$ emission-line luminosity scales appropriately with the UV continuum luminosity. For the following analysis, we do not correct for extinction due to dust and any contamination by the host galaxy; the impact of these, thought to be small, will be quantified in a later study. In Figure~\ref{lumlum}, the continuum luminosity, $\lambda L_\lambda$ at 3000 \AA, is plotted against the emission-line luminosity of H$\alpha$. Our data (as shown by the red points) spans two decades in luminosity and exhibits a clear correspondence between continuum and line emission. Based on our AGN sample, we determine the best-fit linear relation to be $\log (\lambda L_{3000}) = (0.82\pm0.08) \times \log L_{\rm H\alpha} + (9.31\pm3.47)$. For the linear fitting, we adopt a {\tt FITEXY} method \citep{1992nrca.book.....P,2012ApJS..203....6P}. The level of dispersion of the data about this fit is 0.20 dex that will contribute to the dispersion in the final mass estimates.

\begin{deluxetable*}{rlrrlllllllll}
\tabletypesize{\scriptsize}
\tablecaption{Catalog of emission-line properties and black hole masses\label{catalg}}
\tablewidth{0pt}
\tablehead{& & & & & \multicolumn{2}{c}{$\log L$ [erg s$^{-1}$]} && \multicolumn{2}{c}{$\log {\rm FWHM}$ [km s$^{-1}$]} && \multicolumn{2}{c}{$\log M_{\rm BH}$ [$M_\odot$]}\\
\noalign{\smallskip}\cline{6-7}\cline{9-10}\cline{12-13}\noalign{\smallskip}
\colhead{CID\tablenotemark{a}} & \colhead{Field} & \colhead{RA [J2000]} & \colhead{DEC [J2000]} & \colhead{$z$} & \colhead{H$\alpha$} & \colhead{3000 \AA} && \colhead{H$\alpha$} & \colhead{Mg~II} && \colhead{H$\alpha$} & \colhead{Mg~II}}
\startdata
 178 & COSMOS & 149.58521 &  $+$2.05114 & 1.350 & 43.76$\pm$0.03 & 45.35$\pm$0.03 && 3.685$\pm$0.013 & 3.811$\pm$0.079 && 8.68$\pm$0.03 & 8.78$\pm$0.16\\
5275 & COSMOS & 149.59021 &  $+$2.77450 & 1.400 & 44.10$\pm$0.13 & 45.50$\pm$0.02 && 3.835$\pm$0.028 & 3.973$\pm$0.072 && 9.18$\pm$0.09 & 9.18$\pm$0.14\\
 322 & COSMOS & 149.62421 &  $+$2.18067 & 1.190 & 43.16$\pm$0.14 & 45.02$\pm$0.02 && 3.440$\pm$0.044 & 3.581$\pm$0.081 && 7.84$\pm$0.12 & 8.17$\pm$0.16\\
 192 & COSMOS & 149.66358 &  $+$2.08522 & 1.220 & 43.24$\pm$0.14 & 45.05$\pm$0.02 && 3.339$\pm$0.075 & 3.490$\pm$0.035 && 7.68$\pm$0.17 & 8.00$\pm$0.07\\
 157 & COSMOS & 149.67512 &  $+$1.98275 & 1.330 & 42.98$\pm$0.13 & 44.84$\pm$0.03 && 3.701$\pm$0.040 & 3.632$\pm$0.017 && 8.28$\pm$0.11 & 8.19$\pm$0.04\\
\enddata
\tablenotetext{a}{The {\it Chandra} obs. ID.}
\end{deluxetable*}

We can compare our data set to more luminous quasars from the SDSS. In particular, we identify 327 quasars from the SDSS sample in \citet{2011ApJS..194...45S} with $0.35 < z < 0.40$, that cover a similar luminosity range as our high redshift sample. These were selected from 1178 quasars at $0.35 < z < 0.40$ based on high-quality data determined by adopting the following criteria: error in $\log (M_{\rm BH} / M_\odot) < 0.5$, $\log ({\rm FWHM}_{\rm H\alpha} / {\rm km \ s^{-1}}) > 2.9$, and $\log ({\rm FWHM}_{\rm Mg\!~II} / {\rm km \ s^{-1}}) > 2.9$. In addition, we include data from a recent study by \citet{2012ApJ...753..125S} that provides the emission line properties including H$\alpha$ line of high-luminosity quasars from the SDSS with $1.5 < z < 2.2$. These samples are added to our data shown Figure~\ref{lumlum}. We clearly see that SDSS quasars fall along the $\lambda L_{3000}$-$L_{\rm H\alpha}$ relation as established above. Furthermore, the SDSS quasars have similar dispersion at both low-$z$ and high-$z$ samples to our sample, $\sigma = 0.20$ and $\sigma = 0.15$, respectively. We highlight that our AGN sample nicely extends such comparisons between continuum luminosity and line emission at higher redshifts and to lower luminosities. We are able to effectively establish a wider dynamic range, not present in the high-$z$ SDSS sample due to the limited luminosity range around $\log (\lambda L_{3000}) \sim 46.2$. By merging all three samples, we find the following relation based on a linear fit: $\log (\lambda L_{3000}) = (0.96\pm0.01) \times \log L_{\rm H\alpha} + (3.00\pm0.53)$.

While there is very good agreement between the UV continuum and emission line luminosity, there is a small difference that may impact, even slightly, our comparison of the masses. Based on the FMOS sample, the mean ratio $\langle \log (\lambda L_{3000}/L_{\rm H\alpha}) \rangle$ is $1.54\pm0.03$, slightly higher than that found for both the low-$z$ and high-$z$ SDSS quasar samples mentioned above; $\langle \log (\lambda L_{3000}/L_{\rm H\alpha}) \rangle = 1.40\pm0.01$ and $\langle \log (\lambda L_{3000}/L_{\rm H\alpha})\rangle = 1.36\pm0.02$, respectively. This can be seen in the inset histogram in Figure~\ref{lumlum}. If this was due to the effect of dust extinction, one would find the opposite trend with reduced UV continuum relative to the H$\alpha$ line emission. There may be other explanations such as an underlying SED that may be different for X-ray selected samples \citep{2012ApJ...759....6E,2012arXiv1210.3044H} and has an impact on the response seen in photoionized gas, or the effect of the host galaxy on the aperture corrections that differs in each band. While this issue is of importance, we reserve a detailed investigation to subsequent work since it is beyond the scope of this paper to adequately demonstrate such effects. Here, we are primarily concerned with the magnitude of a luminosity offset and whether it contributes to an offset between the masses. With black hole mass scaling with the square root of the luminosity (see below), the offset in luminosity, as determined above, amounts to a very small offset in $\log (M_{\rm BH}/M_\odot)$ of 0.07.

\subsection{Velocity widths}

A second pillar for the use of \ion{Mg}{2} as a black hole mass indicator is that the emitting-line gas is located essentially within the same clouds that emit Balmer emission. While there are claims that this is the case by comparing the velocity profile of \ion{Mg}{2} with H$\beta$ \citep[e.g.,][]{2002MNRAS.337..109M,2011ApJS..194...45S}, there are reported differences and trends that are not well understood \citep{2009ApJ...707.1334W,2012ApJ...753..125S,2012MNRAS.427.3081T}. For example, \ion{Mg}{2} tends to be narrower than H$\beta$ with a difference significantly larger at higher velocity widths \citep[see Figure~2 of][]{2009ApJ...707.1334W}.

\begin{figure}
\epsscale{1.1}
\plotone{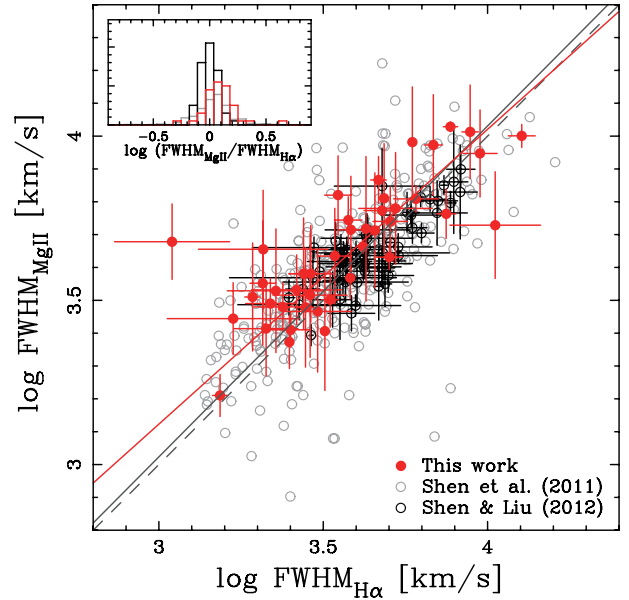}
\caption{Comparison between the FWHM of H$\alpha$ with that of \ion{Mg}{2}. Red, gray, and black circles are the same as those in Figure~\ref{lumlum}. A linear fit to both our data only (red line) and the joint sample (black line) is very similar to a one-to-one relation indicated by the dashed line. The distribution of $\log ({\rm FWHM}_{\rm Mg\!~II} / {\rm FWHM}_{\rm H\alpha})$ is also shown in the sub panel.}
\label{fwhfwh}
\end{figure}

Our aim here is to compare the FWHM of the \ion{Mg}{2} and H$\alpha$ emission lines using a sample not yet explored, namely the moderate-luminosity AGNs at high-redshifts in survey fields such as COSMOS. In Figure~\ref{fwhfwh}, we plot the emission-line velocity width between H$\alpha$ and \ion{Mg}{2} emission lines. Based on the FMOS sample only, a positive linear correlation is seen between the velocity width of the two emission lines with the mean ratio of $\langle \log ({\rm FWHM}_{\rm Mg\!~II} / {\rm FWHM}_{\rm H\alpha}) \rangle = 0.089\pm0.022$ ($\sigma = 0.143$). Our data is in very good agreement with those from the SDSS, i.e., $\langle \log ({\rm FWHM}_{\rm Mg\!~II} / {\rm FWHM}_{\rm H\alpha}) \rangle = 0.033$ ($\sigma = 0.176$) for the low-$z$ sample from \citet{2011ApJS..194...45S}, $\langle \log ({\rm FWHM}_{\rm Mg\!~II} / {\rm FWHM}_{\rm H\alpha}) \rangle = 0.00$ ($\sigma = 0.076$) for the high-$z$ sample from \citet{2012ApJ...753..125S},  and recent FMOS results from SXDS \citep[see][]{2012ApJ...761..143N}. Based on a linear fit to the data shown in Figure~\ref{fwhfwh}, we measure a slope that is consistent with unity; $0.898\pm0.132$ for FMOS only and $1.001\pm0.001$ for FMOS$+$SDSS, and that cannot substantiate the claim by  \citet{2009ApJ...707.1334W} for a shallower value. These results are supportive of a scenario where the \ion{Mg}{2} and the H$\alpha$ emitting regions are essentially co-spatial with respect to the central ionizing source.

There are a few noticeable outliers well outside the dispersion of the sample. These objects then appear as outliers when comparing their masses based on different lines in the next section. Upon inspection, we find that these are the result of the FMOS spectra having low S/N. In some cases, there may be a fit to the H$\alpha$ emission line, based on different parameter constraints, that is equally acceptable to the original fit as assessed by a chi-square goodness of fit and has a velocity width in closer agreement with \ion{Mg}{2}.  Although, we refrain from such selective fitting in order to present results that may be obtained from using similar fitting algorithms on larger data sets where such close inspection is not feasible.

\subsection{Virial single-epoch masses}

We now calculate black hole masses ($M_{\rm BH}$) based on our single-epoch spectra using both (i) $L_{3000}$ and FWHM$_{\rm Mg\!~II}$, and (ii) $L_{\rm H\alpha}$ and FWHM$_{\rm H\alpha}$. This calculation can be expressed as follows:
\begin{equation}
\log \left(\frac{M_{\rm BH}}{M_\odot}\right) = a + b \log \left(\frac{\lambda L_\lambda \ {\rm or} \ L_{\rm line}}{10^{44} \ {\rm ergs \ s}^{-1}} \right) + c \log \left(\frac{{\rm FWHM}}{{\rm km \ s}^{-1}} \right).
\end{equation}
We explicitly use the recipes provided by \citet{2005ApJ...630..122G} and \citet{2004MNRAS.352.1390M} for the cases of H$\alpha$ and \ion{Mg}{2} lines, respectively:
\begin{eqnarray*}
(a,b,c) &=& (1.221,0.550,2.060) \ \ \ \ \ \ \ \ \ \ {\rm for \ H\alpha},\\
(a,b,c) &=& (0.505,0.620,2.000) \ \ \ \ \ \ \ \ \ \ {\rm for \ Mg\!~II}.
\end{eqnarray*}
We note that the calibration of the relation for \ion{Mg}{2} has been carried out by many studies \citep[e.g.,][]{2006ApJ...641..689V,2008ApJ...673..703M} and there are known differences between them.

\begin{figure}
\epsscale{1.1}
\plotone{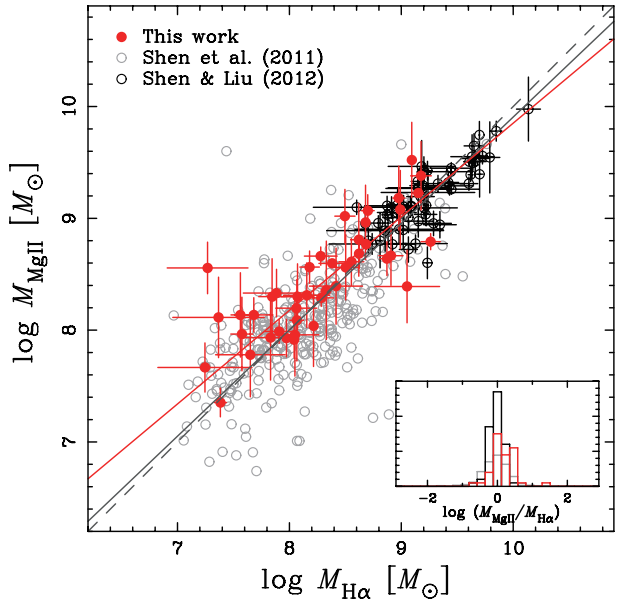}
\caption{The comparison of the black hole mass estimated by H$\alpha$ line with that by \ion{Mg}{2} line. Red, gray, and black circles are the same as those in Figure~\ref{lumlum}. The fit to both our data (red line) and the joint sample (black line) is nearly equivalent to a one-to-one relation (dashed line). The distribution of $\log (M_{\rm Mg\!~II} / M_{\rm H\alpha})$ is also shown in the sub panel.}
\label{masmas}
\end{figure}

In Figure~\ref{masmas}, we show $M_{\rm BH}$ for our sample derived from H$\alpha$ and \ion{Mg}{2}. Our sample spans a range of $7.2 \la \log (M_{\rm BH}/M_\odot) \la 9.5$ consistent with that reported by previous studies of type 1 AGNs in COSMOS \citep{2010ApJ...708..137M,2011ApJ...733...60T} and is complementary to the higher-$L$ quasar sample at similar redshifts with $\log (M_{\rm BH}/M_\odot) \ga 9$ \citep{2012ApJ...753..125S}. We find the average (dispersion) in the black hole mass ratio of $\langle \log (M_{\rm Mg\!~II}/M_{\rm H\alpha}) \rangle$ is 0.17 ($\sigma = 0.32$) for our FMOS sample. These results are similar to that determined from the SDSS sample; the average (dispersion) in the black hole mass ratio is $-0.05$ ($\sigma = 0.39$) at low $z$ and $-0.03$ ($\sigma = 0.20$) at high $z$. While an offset of 0.17 dex is seen in the FMOS sample, we conclude that the recipes established using local relations give consistent results between \ion{Mg}{2}- and H$\alpha$-based estimates.

\section{Conclusions}

We have investigated the emission line properties of AGNs in COSMOS and ECDF-S to establish whether \ion{Mg}{2} and H$\alpha$ provide comparable estimates of their black hole mass. This study is the first attempt to do so for AGNs of moderate luminosity, hence lower black hole mass ($10^7 M_\odot < M_{\rm BH} < 10^9 M_\odot$), at high redshift that complements studies of more luminous quasars. Our results clearly show that the velocity profiles of \ion{Mg}{2} and H$\alpha$ are very similar when characterized by FWHM and the relation between continuum luminosity and line luminosity is tight. We then find that virial black hole masses based on \ion{Mg}{2} and H$\alpha$ have very similar values and a level of dispersion ($\sigma \sim 0.3$) comparable to luminous quasars from SDSS. It is important to keep in mind that these results pertain to specific calibrations \citep{2004MNRAS.352.1390M,2005ApJ...630..122G} for estimating black hole mass. The use of other recipes, such as provided by \citet{2008ApJ...673..703M}, will show a discrepancy larger than seen here. To conclude, the locally-calibrated recipes for black holes masses using \ion{Mg}{2} and H$\alpha$ are applicable for fainter AGN samples at high redshift. These results further support a lack of evolution in the physical properties of the broad line region in terms of quantities such as $\alpha_{\rm ox}$ \citep[e.g.,][]{2006AJ....131.2826S,2009ApJ...690..644G}, emission-line strengths \citep[e.g.,][]{2006NewAR..50..665F}, and the inferred metallicities \citep{2001A&A...372L...5M,2006A&A...447..157N,2011A&A...527A.100M}.

As a final word of caution, such estimates of black hole mass are likely to have inherent dispersion as discussed above and systematic uncertainties that are not yet well understood. For instance, recipes for estimating black hole mass depend on the assumption that the gas is purely in virial motion. This is unlikely to be true for all cases since both outflows and inflows are common in AGNs. Even so, there is evidence that the virial product of mass and luminosity (as a proxy for the radius to the BLR) is a useful probe of the central gravitational potential. In the very least, it is important to establish the level of dispersion in such relations since observed trends usually rely on offsets comparable to the dispersion such as the redshift evolution of the relation between black holes and their host galaxies \citep[e.g.,][]{2006ApJ...640..114P,2010ApJ...708..137M,2012arXiv1212.2999S}.

\acknowledgments

We thank Kentaro Aoki and Naoyuki Tamura for their invaluable assistance during our Subaru/FMOS observations. K.M. acknowledges financial support from the Japan Society for the Promotion of Science (JSPS). Data analysis were in part carried out on common-use data analysis computer system at the Astronomy Data Center, ADC, of the National Astronomical Observatory of Japan (NAOJ).

\end{document}